# Decision making based on optical excitation transfer via near-field interactions between quantum dots

Makoto Naruse,[1,a)] Wataru Nomura,[2] Masashi Aono,[3,4] Motoichi Ohtsu,[2] Yannick Sonnefraud,[5,6] Aurélien Drezet,[5,6] Serge Huant,[5,6] and Song-Ju Kim[7]

1 Photonic Network Research Institute, National Institute of Information and Communications Technology, 4-2-1 Nukui-kita, Koganei, Tokyo 184-8795, Japan

2 Department of Electrical Engineering and Information Systems, Graduate School of Engineering, The University of Tokyo, 2-11-16 Yayoi, Bunkyo-ku, Tokyo 113-8656, Japan

3 Earth-Life Science Institute, Tokyo Institute of Technology, 2-12-1 Ookayama, Meguru-ku, Tokyo 152-8550, Japan

4 PRESTO, Japan Science and Technology Agency, 4-1-8 Honcho, Kawaguchi-shi, Saitama 332-0012, Japan

5 Université Grenoble Alpes, Inst. NEEL, F-38000 Grenoble, France

6 CNRS, Inst. NEEL, F-38042 Grenoble, France

7 WPI Center for Materials Nanoarchitectonics, National Institute for Materials Science, 1-1 Namiki, Tsukuba, Ibaraki 305-0044, Japan

a) Electronic mail: naruse@nict.go.jp

**Abstract:** Optical near-field interactions between nanostructured matter, such as quantum dots, result in unidirectional optical excitation transfer when energy dissipation is induced. This results in versatile spatiotemporal dynamics of the optical excitation, which can be controlled by engineering the dissipation processes and exploited to realize intelligent capabilities such as solution searching and decision making. Here we experimentally demonstrate the ability to solve a decision making problem on the basis of optical excitation transfer via near-field interactions by using colloidal quantum dots of different sizes, formed on a geometry-controlled substrate. We characterize the energy transfer behavior due to multiple control light patterns and experimentally demonstrate the ability to solve the multi-armed bandit problem. Our work makes a decisive step towards the practical design of nanophotonic systems capable of efficient decision making, one of the most important intellectual attributes of the human brain.

## I. INTRODUCTION

Optical excitation transfer between quantum nanostructures, such as quantum dots (QDs), is one of the most unique and valuable physical processes in nanophotonics.[1-3] When QDs share common resonant energy levels mediated by optical near-field interactions, optical energy is transferred from smaller QDs to larger ones. Optical near-field interactions even allow transitions that are conventionally dipole forbidden thanks to the localized property of the optical near-fields.[1,4] By exploiting such attributes, optical energy transfer has been applied to a wide range of applications, including energy concentration,[3] logic circuits,[5,6] and sensing.[7] Furthermore, the possibility of *intelligent abilities* by utilizing optical energy transfer has been demonstrated theoretically; for instance, the ability to solve a constraint satisfaction problem (CSP) has been demonstrated[8] as well as satisfiability problem (SAT),[9] decision making problems.[10] In addition, novel architectures based on recent optical technologies to accomplish computing capability have been presented.[11-13]

In this paper, we experimentally demonstrate a decision making principle based on optical excitation transfer between two kinds of different-sized colloidal QDs formed on a geometry-engineered substrate. What we particularly discover in our investigation of intelligence-related applications of optical excitation transfer is that, until the energy dissipation is induced, the optical excitation exists somewhere across the resonant energy levels in a nonlocalized manner, whereas it is transferred to a particular destination quantum dot when energy is dissipated. Moreover, the probability of which destination quantum dot, out of many, is chosen depends on the surrounding environment. Specifically, if the destination energy level(s) is occupied by another excitation(s), resulting in what is called state-filling effects,[14-17] the input optical excitation is more likely to be transferred to other unoccupied energy levels. At the same time, it should also be noted that the probability of the excitation going to a dot irradiated with control light is not perfectly zero if the state-filling does not perfectly inhibit the transition.[8-10] With these fundamental mechanisms, our work opens the way to the practical design of nanosystems capable of decision making, one of the main attributes of human intelligence.

## II. QUANTUM DOT-BASED DECISION MAKER

### A. Architecture

Consider the particular case of a decision making problem in which a *player* should make a quick and accurate decision in choosing, from many available ones, a *slot machine* that has the highest probability of paying out a reward so that the player can get as much reward as possible. To accomplish this, the player should test and evaluate which machine is the best; however, too much testing or exploration to search for the best machine may result in a significant loss. Moreover, the best machine may change with time. In other words, there is a trade-off between *exploration* and *exploitation*, referred to as the *exploration–exploitation dilemma*.[18] Such a problem, called the *multi-armed bandit problem* (BP), is the foundation of many important applications in information and communication technologies, such as frequency assignment in wireless communication networks,[19,20] web-content optimization,[21] Monte Carlo tree searches,[22] etc. Decision making is also an important issue in neurosciences, and even clinical implications have been discussed.[23] For the most simple case, while still preserving the essence of the problem, we restrict ourselves to choosing the best of two slot machines. We refer to the two machines as **slot machine L** and **slot machine R** hereafter.

In Ref. 10, Kim *et al.* presented a quantum dot-based decision maker (QDM) consisting of five QDs networked via optical near-field interactions (Fig. 1). Here, we briefly outline the basic mechanism of the QDM.[10] An optical excitation generated at the smaller-sized QD labeled **QD$_S$** in Fig. 1 can be transferred to either one of the large-sized QDs (labeled **QD$_{LL}$** and **QD$_{LR}$**) through inter-dot optical near-field interactions. This means that the photon radiation from the medium-sized QDs (denoted by **QD$_{ML}$** and **QD$_{MR}$**) is negligible. However, when the large-sized QDs, **QD$_{LL}$** and **QD$_{LR}$**, are irradiated with control light that induces state-filling effects, the probabilities of photon radiation from **QD$_{ML}$** and **QD$_{MR}$** do appear. Such a phenomenon can be regarded as if the input photon, generated at **QD$_S$**, is subjected to a *tug-of-war,* being pulled by the larger-size QDs on the left- and right-hand sides, meaning that when one side is pulled, the other side is immediately pushed. Such a *nonlocal correlation*, in the sense that the state-filling induced at one dot immediately alters the dynamics of the entire system, has been shown to enhance the performance in solving the BP.[24] Note also that fluctuation is essential in order to realize an "exploration" ability; that is to say, the probability of the excitation going to a dot irradiated with control light is not perfectly zero if the state-filling does not perfectly inhibit the transition, as mentioned earlier.

Here we experimentally demonstrate such a mechanism by using combinations of different-sized QDs irradiated by control light patterns, evaluate the basic characteristics of optical excitation transfer, and finally demonstrate the ability to solve decision making problems in dynamically changing environments. Note that the probabilistic nature of a photon plays a key role, as a first step toward future integrated decision making machines, and here we employ an ensemble of many quantum dots to experimentally realize optical energy transfer. The probabilistic nature is emulated by an electrical controller. In other words, the probabilistic attributes are simulated by the host controller while its probabilities are determined by the observation of optical excitation transfer involving a large number of QDs. Nevertheless, we believe that this is an important first experimental demonstration that optical near-field-mediated energy transfer can solve decision making problems.

The idea of a tug-of-war is represented by the notion of an *"intensity adjuster"* (IA),[10] which physically corresponds to a mechanism that modifies energy transfer patterns in the system in the following manner.

1. The IA's value of "zero" indicates no imbalance of optical excitation transfer in the system. That is, control light of the same intensity is applied to *both* of the lowest energy level of **QD$_{LL}$** and **QD$_{LR}$**, which are respectively denoted by LL$_1$ and LR$_1$ in Fig. 1.

2. For decision making, we focus our attention on the photon radiation from the medium-size dots (**QD$_{ML}$** and **QD$_{MR}$**). The decision to choose the designated slot machine is made based on information about from which dot a photon is observed. More specifically, if a photon is observed from **QD$_{ML}$**, that is to say, if radiation is observed from the energy level ML$_1$ denoted in Fig. 1, the decision is made to choose the **slot machine L**, whereas if a photon is observed from **QD$_{MR}$**, that is to say, if radiation is observed from the energy level MR$_1$ in Fig. 1, the decision is made to choose the **slot machine R**.

3. If a reward is successfully dispensed from the selected slot machine, the IA is "moved" in the direction of the selected machine. More specifically, when the chosen machine is the **slot machine L** and a reward is successfully obtained, more control light is applied to LL$_1$, and simultaneously, less light is applied to LR$_1$. Note that the slot machines are "external" systems from the QD-based decision maker. In contrast, if no reward is dispensed, the IA is moved in the opposite direction from the machine chosen. That is to say, when the chosen machine is the **slot machine L**, and a reward is not obtained, less light is applied to LL$_1$, and more light is applied to LR$_1$.

It was demonstrated that the resultant decision making performance exhibits even better performance than that of the best conventionally known algorithm called Softmax[18] as shown in Ref. 10.

**B. Experimental devices**

In the experiment, instead of using three types of dots, we employed two types of dots, referred to as **QD$_S$** and **QD$_L$**, arranged in the configuration **QD$_L$-QD$_S$-QD$_L$**, as schematically shown in Fig. 2(a). As described below, there were many **QD$_S$**s and **QD$_L$**s in the experimental device. In Fig. 2(a), three **QD$_S$**s are indicated. Their energy levels are indicated by $S_1$, $LL_1$, $LL_2$, $LR_1$, and $LR_2$, as shown in Fig. 2(a). (Note that "**QD$_L$**" in Fig. 2(a) corresponds to the "medium"-sized dot (**QD$_M$**) in the former five dot system shown in Fig. 1.) Furthermore, the experimental device, shown below, consists of a large number of smaller and larger dots arranged in a shape-engineered substrate made by lithography. The design of the experimental device is shown in Fig. 2(b). The smaller-size CdSe/ZnS core-shell QDs (**QD$_S$**s) are distributed in two 1 μm-wide lanes joined end-to-end, depicted by the dotted area in Fig. 2(b), whereas the larger-size ones (**QD$_L$**s) occupied square-shaped areas, with a side length of 3 μm, attached to the ends of the two lanes. The distance from the center of the **QD$_S$** area, indicated by $P_S$, and the centers of the two **QD$_L$** areas, denoted by $P_L$ and $P_R$, were designed to be 7.5 μm.

Such dimensions are larger than the diffraction limit of the light used in the experiment, which in fact sacrifices one of the fundamental benefits of optical excitation transfer via near-field interactions, namely, the possibility to operate at a scale below the diffraction limit. In this paper, however, the primary focus is a first experimental demonstration of the decision making function, and we consider that a macro-scale setup suffices, as long as the underlying essential principles are preserved. We have previously confirmed optical excitation transfer mediated by near-field interactions based on these distributed colloidal quantum dots and have demonstrated optical excitation transfers over distances as long as 10 μm.[25] For these reasons, the fabricated experimental device was based on a large number of QDs arranged in the architecture shown in Fig. 2(b).

The input light is radiated onto the area around the position $P_S$ in Fig. 2(b), where smaller QDs are located, corresponding to the generation of an optical excitation at the energy level $S_1$ in the model shown in Fig. 2(a). On the other hand, the control light for inducing state-filling effects in the energy levels $LL_1$ and $LR_1$ is radiated onto the areas occupied by the **QD$_L$**s around the positions $P_L$ and $P_R$, as schematically shown by the dashed circles labeled CL and CR in Fig. 2(b). Since the two regions CL and CR are separated by a distance larger than the wavelength, experimentally we are able to address each area with a diffraction-limited beam (generated by a computer generated hologram (CGH), described below). The areas CL and CR respectively correspond to **QD$_{LL}$** and **QD$_{LR}$**, and thus CL and CR are also indicated in Fig. 2(a).

We fabricated QD samples using the procedure where a lithography and a lift-off technique were repeated twice. We used commercially available colloidal CdSe/ZnS core-shell QDs (manufactured by Quantum Design, Inc.). The core-diameters of **QD$_S$** and **QD$_L$** were 2.5 nm and 3.2 nm, and the first excited states (schematically shown in Fig. 2(a)) were $S_1$ = 2.36 eV (525 nm) and $LR_1$ = $LL_1$ = 2.11 eV (588 nm), respectively. With such a combination of QDs, the energy level $S_1$ is resonant with the second excited states of the larger dots, that is, $LR_2$ and $LL_2$.[25] (1) First, by using e-beam lithography on a silica substrate coated with resist (ZEP-520A, Zeon Corp.) with a thickness of around 100 nm, (2) the two areas in which **QD$_L$**s were to be deposited were formed. (3) Next, the **QD$_L$** solution was dripped onto the substrate and was allowed to dry naturally at room temperature. (4) By removing the resist, we obtained the **QD$_L$** structure on the substrate. (5) Next, resist was formed on top of the structure, and (6) using e-beam lithography once again, the structures on which **QD$_S$**s were to be deposited were formed by aligning their positions with the former **QD$_L$** structures. (7) After developing the resist, we dripped the **QD$_S$** solution onto the substrate. (8) Finally, the resist was removed to obtain the final structure. Figure 2(c) shows a fluorescence microscope image obtained when the fabricated sample was excited with a mercury-vapor lamp. The irradiation ultraviolet light from a mercury-vapor lamp efficiently excites CdSe/ZnS core-shell QDs, allowing us to observe the entire device structure in the visible region. The **QD$_S$** and **QD$_L$** structures were respectively observed as green and yellow colors. Due mainly to alignment precision errors in the liftoff processes, the distances between the center of the **QD$_S$** area and the two **QD$_L$** areas (left-hand and right-hand sides) were 2.01 μm and 3.85 μm, which are respectively indicated by dotted and dashed lines.

The radiation of the output light is correlated with the radiation from the smaller dots (**QD$_S$**s) located around the edge of the lanes or the area intersecting with **QD$_L$**s, which are denoted by DL and DR as indicated by the circles in Fig. 2(b). The higher the occupation probability of the first excited state of **QD$_L$**, the more the energy transfer to the designated **QD$_L$** is prohibited, leading to increased radiation from **QD$_S$**s located in their proximity. Such a mechanism exhibits the same character as that theoretically demonstrated in Ref. 10 in the sense that more radiation from the output dot (**QD$_M$**) results from more state filling in its neighboring **QD$_L$**. The areas DL and DR respectively correspond to **QD$_S$**s located close to **QD$_{LL}$** and **QD$_{LR}$** in Fig. 2(a), and thus DL and DR are also indicated in Fig. 2(a).

## C. Experimental systems

After the experimental device fabrication by using lift-off techniques and two kinds of colloidal quantum dots, the next important concern in implementing decision making functions is how to apply the two control light beams to irradiate the areas CL and CR. For this purpose, we introduced a phase-only spatial light modulator (SLM), based on liquid crystal on silicon (LCOS) technology (LCOS-SLM X10468-01, Hamamatsu Co. Ltd.). It has 792 × 600 pixels with a pixel pitch of 20 μm, corresponding to a maximum spatial frequency of 25 lp/mm, designed for the wavelength region between 400 nm and 700 nm, and each pixel can modulate the phase in 256 grayscale levels. The diffraction efficiency of the LCOS-SLM is about 30 % for a spatial frequency of 25 lp/mm.

Continuous-wave (CW) light at a wavelength of 589 nm produced by a diode-pumped solid-state laser (Shaghai Dream Laser Inc., MGL-W-589-1W) was incident normally on the SLM, and the reflected light was collected by an objective lens with a numerical aperture (NA) of 0.40 and supplied to the sample. The experimental setup is schematically shown in Fig. 3(a). Computer-generated holograms (CGHs) with a size of 256 × 256 pixels, with 256 grayscale levels, were displayed on the SLM. The CGHs were designed so that their Fourier transform yielded two points corresponding to the first diffraction orders with intensities that can be independently adjusted. Let $G_L$ and $G_R$ denote the grayscale levels of the respective areas. For example, with the combination $(G_L, G_R)=(128,128)$, the control light beams radiated onto the areas CL and CR have equal intensities, whereas the combination $(G_L, G_R)=(255,1)$ means that the area CL is irradiated much more intensely than CR is. We prepared multiple CGHs beforehand and switched between them during the characterization of the device and the decision making demonstration shown below.

The optical excitation in the smaller QDs is generated by incident light focused on the position around $P_S$ from a Ti:sapphire laser (Coherent Inc., Mira900) with a wavelength of 360 nm, a pulse width of 2 ps, and a repetition frequency of 80 MHz through the objective lens that the above control light also passes through. The radiation from the device was observed from the back surface through an objective lens with an NA of 0.55, followed by a bandpass filter with a pass wavelength of 540 nm ± 5 nm, and was captured by an electron multiplying CCD camera (Hamamatsu Co. Ltd., ImagEM C9100-13H) which acquired images with a 16-bit grayscale and 512 × 512 pixels with a resolution of 0.37 μm/pixel. Figure 3(b) shows the ratio of the control light intensity obtained by evaluating the average pixel values in the areas CR and CL shown in Fig. 2(c). We extracted two regions-of-interest (ROIs) consisting of 3 × 3 pixels from the image captured by the EMCCD camera, which corresponds to a 1.1 μm × 1.1 μm area in the QD device. The $G_R$ value of the CGH was adjusted while maintaining $G_L + G_R = 256$. As shown in Fig. 3(b), as the $G_R$ value increased, the light intensity at CR, denoted by circular marks, increased, whereas that at CL, denoted by square marks, decreased.

We then evaluated the basic character of optical excitation transfer. The square and circular marks in Fig. 3(c) respectively indicate the average pixel values corresponding to the areas DL and DR, obtained by extracting two 3 × 3 pixel ROIs, as a function of the ratio of the control light intensity incident on CR. The signal levels at DR (circular marks) increased, whereas those at DL (square marks) decreased as the imbalance of the intense control light intensity was shifted from left to right, which is consistent with the theoretical investigation discussed earlier. The absolute values in the areas DL and DR should ideally be symmetric when the same intensity of light is radiated onto CL and CR, but the experimental results shown in Fig. 3(c) did not exhibit such perfect symmetry. Also, the variance of the signal in DR was larger than that in DL; this is because, as shown in Fig. 2(c), the device structure was not perfectly symmetric due to the fabrication process, mainly because of the second liftoff step. In addition, optical misalignment of the overall experimental system may have caused a certain imbalance with respect to both light irradiation and observation. Nevertheless, it should be emphasized that the dependencies of the output signals from DL and DR on the imbalance of the control light controlled by the CGHs followed the "tug-of-war"-type behavior, meaning that the increase of the signal at one side and the decrease at the other side are correlated. This is one critical feature for the decision making mechanism.

Despite the device imperfections described above, one *presumable* reason behind the low-contrast for DL and the high-contrast for DR is as follows. Suppose that the distance between DL (or DR) and $P_S$ is correlated with the likelihood of radiation from DL (or DR). When state-filling light is radiated at CL, radiation from DL is more likely to be induced regardless of the power of the state-filling light radiated at CL, and thus the resulting contrast is low. On the other hand, when state-filling light is radiated at CR, the radiation from DR depends more sensitively on CR, and thus the contrast is high. In other words, due to the large distance between DR and $P_S$, an optical excitation around DR is more strongly affected by the excitations in CR, resulting in high contrast. Note, however, that the above mechanism regarding the contrast is merely speculation, and it will be an interesting topic to examine in a future study.

## III. DECISION MAKING DEMONSTRATIONS

From the nature of optical excitation transfer observed in Fig. 3(c), the probability of observing a photon from either **$QD_{LL}$** or **$QD_{LR}$** is determined by the IA. Thus, an immediate single photon measurement determines the decision of which slot machine to choose, whereas the photon observation probability is engineered by state-filling effects. By utilizing the experimentally observed characteristics of optical energy transfer and by representing its probabilistic nature, in the work described in this paper, we took the following approach rather than directly employing a single photon measurement. Such an approach also takes account of the above-mentioned experimentally unavoidable imbalance in the fabricated device. Moreover, we consider that the demonstration will lead to more sophisticated experimental systems, including single photon measurement systems.

Let the intensities observed at DL and DR be denoted by $I_L$ and $I_R$. These values are then calibrated by using the expressions $\widehat{I_L} = A_L I_L - B_L$ and $\widehat{I_R} = A_R I_R - B_R$, where $A_L$, $A_R$, $B_L$, and $B_R$ are constants. Based on these values, a *threshold* value is defined by

$$T = \frac{\widehat{I_L}}{\widehat{I_L} + \widehat{I_R}}. \tag{1}$$

The threshold $T$ is compared with a random number between 0 and 1 generated at the host computer. If the random number is equal to or larger than $T$, the decision is made to choose the slot machine L, whereas if it is smaller than $T$ the decision is made to choose the slot machine R.

The intensity adjustor (IA) is implemented as follows. The grayscale values for the left- and right-hand sides at cycle $t$ are given by

$$G_L(t) = \text{CGH}_L(\lceil IA(t) \rceil) \tag{2}$$

$$G_R(t) = \text{CGH}_R(\lceil IA(t) \rceil) \tag{3}$$

where $\lceil \ \rceil$ means the floor function, which truncates the decimal part. The functions $\text{CGH}_L(n)$ and $\text{CGH}_R(n)$ specify the grayscale values for CL and CR, respectively, when the truncated IA value is given by $n$. Let the initial IA value be $IA_0$. If the slot machine L yields a reward in cycle $t$, the IA value is updated based on

$$IA(t+1) = -\Delta + IA_0 + \alpha(IA(t) - IA_0) \tag{4}$$

where $\alpha$ is referred to as a forgetting parameter, and $\Delta$ is a constant increment of IA[10], which is assumed to be unity in the experiments. In the case where the machine R gives a reward at the cycle $t$, the update rule is given by

$$IA(t+1) = +\Delta + IA_0 + \alpha(IA(t) - IA_0). \tag{5}$$

In the following decision making demonstration, the truncated integer values of the IA value are assumed to take a natural number -3, -2, -1, 0, 1, 2, or 3. When $IA(t) \geq 4$ or $IA(t) \leq -4$, the truncated IA value is respectively given by $\lceil IA(t) \rceil = +3$ or $\lceil IA(t) \rceil = -3$. The particular $G_L$ and $G_R$ values are specified by seven CGHs, providing the $(G_L, G_R)$ pairs with pixel values (255,1), (148,108), (136,120), (126,130), (118,138), (104,152), and (1,255), which respectively correspond to truncated intensity adjustor values of -3, -2, -1, 0, 1, 2, and 3. Figure 3(d) summarizes the mechanism of the experimental system used to solve the multi-armed bandit problem.

Let the reward probability of the machine L be $P_L = 0.8$ and the reward probability of the machine R be $P_R = 0.2$, where the total probability, $P_L + P_R$, is unity, which is the same condition imposed in the model results discussed in Ref. 10. Furthermore, the reward probability is dynamically changed from time to time; the occurrence of a "change" of the reward probability is not notified to the player. In the experiment, the values of $P_L$ and $P_R$ are swapped every 150 plays. The solid curve in Fig. 4(a) shows the evolution of the "success rate", taking a value between 0 and 1, for 600 consecutive plays. Such 600 plays of the slot machines were repeated 10 times. "Success" means choosing the slot machine with the higher reward probability. The success rate is evaluated by calculating the number of successes divided by the number of repeat cycles. The success rate increases as time evolves. As a result of swapping the reward probabilities, the success rate drops every 150 plays but quickly recovers. Such rapid and accurate adaptability to the external environment demonstrates the success of the method of solving the multi-armed bandit problem considered in this paper. Note that the "correct" machine is not, of course, notified to the player; the player may know the "winning rate", which is the ratio of getting coins from the chosen slot machine, which fluctuates considerably. Nevertheless, the "correct selection rate" does indeed approach unity after a certain number of cycles.

The dotted curve in Fig. 4(a) shows the evolution of the success rate with the reward probabilities $P_L = 0.6$ and $P_R = 0.4$. Like the former case, the reward probabilities were switched every 150 plays. Since the difference between the reward probability is smaller than in the former case, accurate decision making is difficult, which is manifested by the fact that the dashed curve stays lower than the former solid curve. However, what is remarkable is that the success rate again increases as time evolves, and the adaptive behavior is also observed in such a difficult situation. Since the machine with the higher reward probability may not dispense a reward in some trials, the "winning rate", which is the ratio of getting coins from the chosen slot machine, is different from the success rate. Figure 4(b) shows the evolution of the winning rate, where we can observe that it approaches 0.8 and 0.6, denoted by solid and dotted curves, respectively, which is the higher probability of the assumed slot machine's reward probability, indicating that the winning rates approach achievable limits.

Behind such success in selecting a higher-reward machine is the way in which the intensity adjustor mechanism works. Figure 4(c) represents the evolution of the IA value, which evolves toward a smaller value (choosing the **slot machine L** is more likely) and a larger one (choosing the **slot machine R** is more likely), and this is consistent with the given reward probabilities. We should also note that, in the case of $P_L=0.8$ and $P_R=0.2$, the duration required to recover unity (that is, 100%) "correct selection rate" after a sudden environmental change is not constant; the durations are about 50, 30, and 20 cycles. These are correlated with the value of IA. At cycle 150, IA is about -500, whereas at cycle 300, IA is about 400. Such a difference in the value of IA, and its associated performance differences, also clearly suggests that our proposed dynamics of IA and the quantum-dot based devices play a key role. In other words, the probabilistic attributes of the IA and the fact that the probability is determined by optical excitation transfer are key. While Fig. 4 demonstrates the dynamics of the proposed decision making mechanism, Fig. 3(c) confirms one important internal process involving optical excitation transfer. In addition, the variance in the recovery time indicates that a more intelligent architecture for the IA may be possible. These results demonstrate the accurate and adaptive decision making ability of the decision maker based on quantum dots mediated by optical near-field interactions.

Finally, we make a remark regarding possible processing speeds and minimum power requirements of this device. The speed of the optical excitation transfer between quantum dots is on the order of 100 ps, and the radiation from QDs is on nanosecond order; in principle, therefore, the operating speed of the QD device itself can be as fast as 100 MHz to GHz order.[26] In the present experimental system, however, the primary bottlenecks are image acquisition (EMCCD camera), and data transfer to the host computer and post processing, which takes about 2 seconds per frame. Such a problem can be resolved by incorporating a fully parallel architecture, such as the one proposed by Ishikawa *et al.* in Ref. 27. The minimum power "dissipation" of optical energy transfer is about $10^4$-times smaller than that of the bit flip energy of electrical devices, for which theoretical[28] and experimental[26] values have been reported. The minimum power "requirement", however, has not been clarified yet, and this will be an important research topic in the future.

## IV. SUMMARY

In summary, we experimentally demonstrated the ability to solve multi-armed bandit problems on the basis of optical excitation transfer via near-field interactions by using different-sized colloidal quantum dots. Small-sized and large-sized QDs were arranged in a geometry-controlled manner by repeating two lithography and lift-off processes. The patterns of optical excitation transfer, from smaller QDs to larger ones, were dynamically reconfigured by modulating the intensity of control light radiated onto the region occupied by the larger QDs to induce state filling, which was experimentally implemented by computer-generated holograms displayed on a spatial light modulator. The basic character of optical excitation transfer was evaluated, and accurate and adaptive decision making was successfully demonstrated. In the future, it would be rewarding to develop genuine sub-wavelength decision makers and to engineer related processes using single photon sources, for instance from nanodiamonds,[29] to establish single-photon decision makers.

## ACKNOWLEDGEMENTS

This work was supported in part by Strategic Information and Communications R&D Promotion Programme (SCOPE) of the Ministry of Internal Affairs and Communications and the Core-to-Core Program, A. Advanced Research Networks from the Japan Society for the Promotion of Science.

**Figure captions**

**Figure 1** Architecture and theoretical analysis of quantum dot-based decision maker (QDM). The QDM is interacting with dynamically changing external environments.

**Figure 2** Quantum dot devices for decision making. (a) Schematic architecture of the experimental device used in the decision making demonstration. (b) Design of the quantum dot-based device. (c) Light emission image from the fabricated quantum dot device.

**Figure 3** Experimental systems for the quantum dot-based decision maker. (a) Schematic diagram of the experimental system. (b) Ratio of the control light intensity radiated onto the areas CR and CL (in Figure 2c) as a function of CGH setup. (c) Signal levels from the areas DR and DL (in Figure 2c) as a function of the imbalance of control light intensities incident on CR and CL. (d) The overall mechanism of the experimental system used in solving the multi-armed bandit problem.

**Figure 4** Decision making demonstration. (a) Correct selection rate, (b) winning rate, and (c) the value of the intensity adjustor (IA).

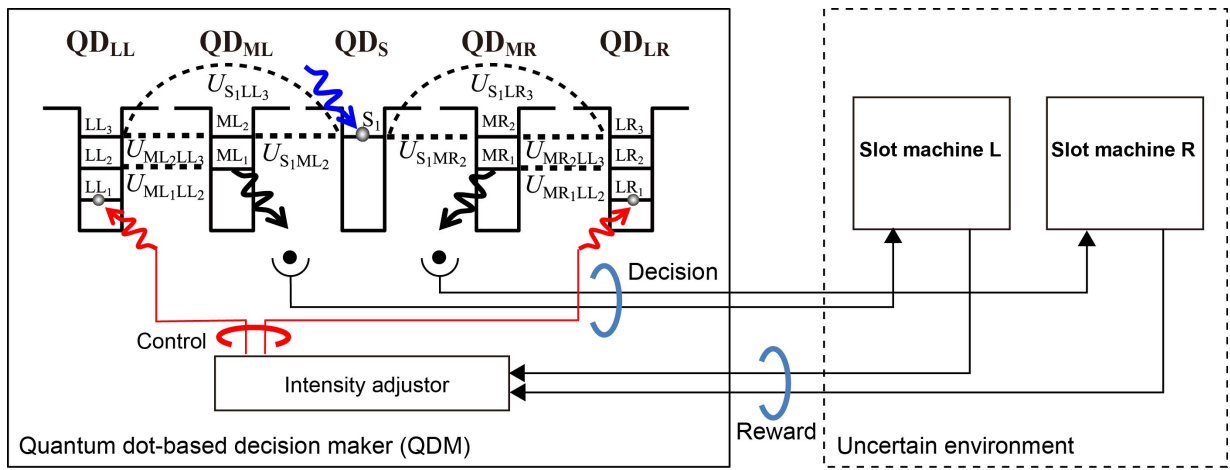

Figure 1

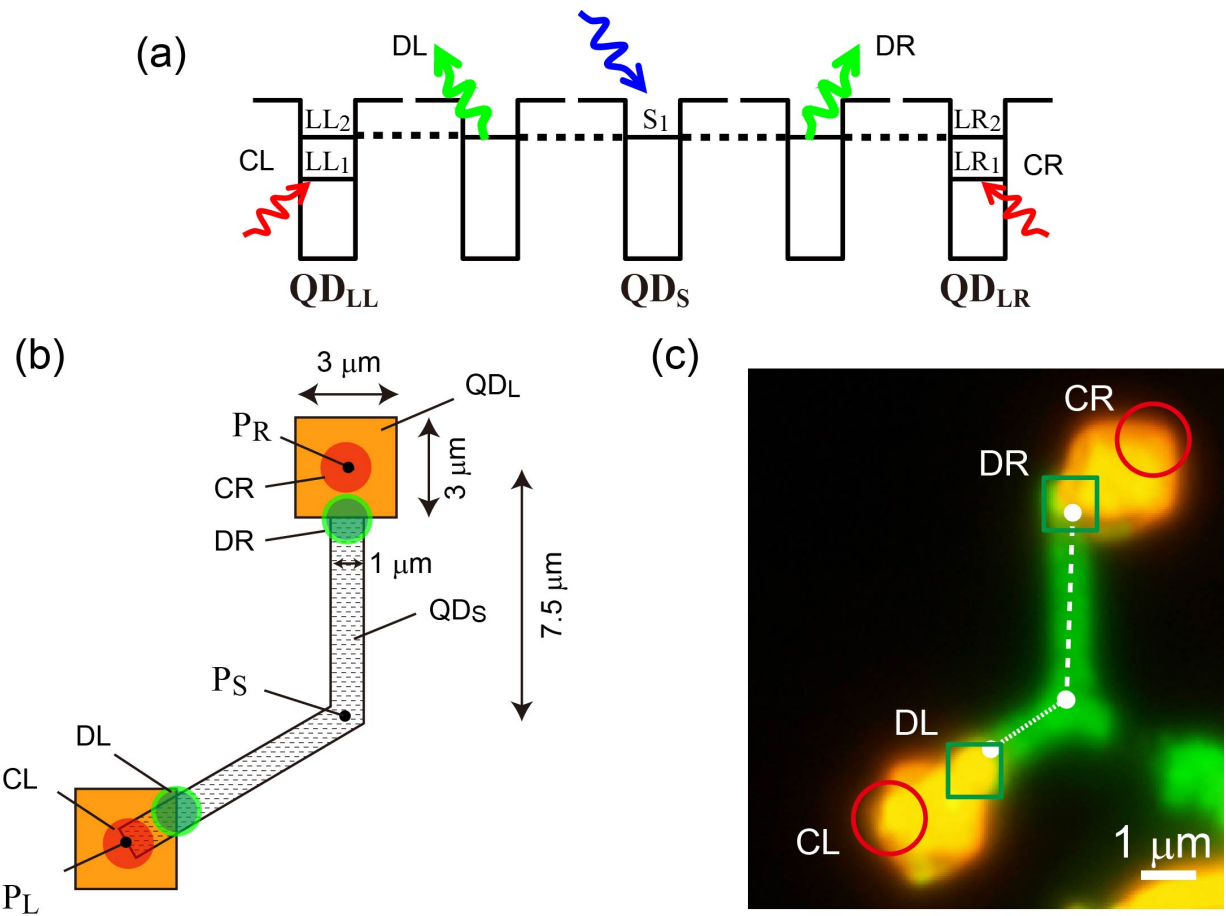

Figure 2

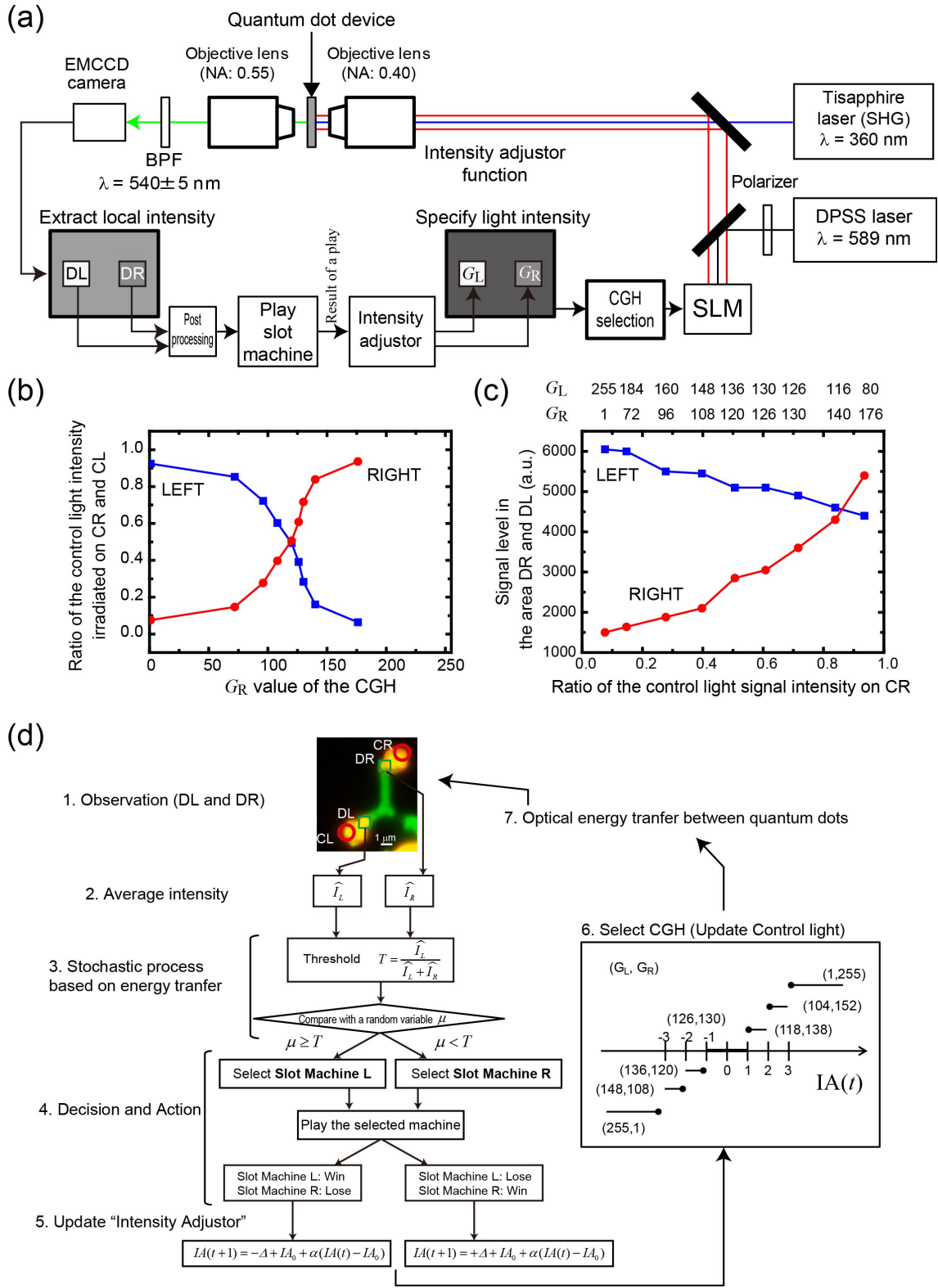

Figure 3

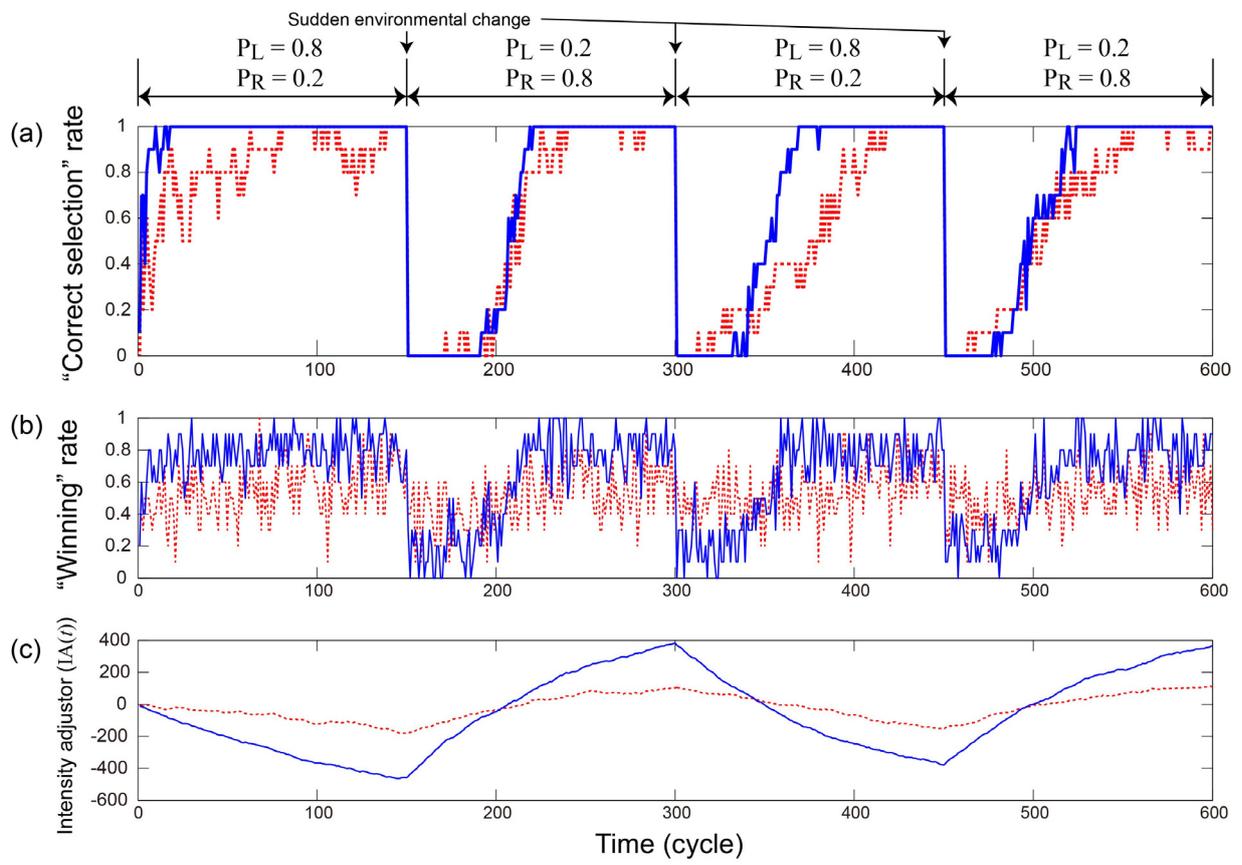

Figure 4